\documentclass[twocolumn,twocolappendix]{openjournal}
\renewcommand{\baselinestretch}{1.075} 
\shortauthors{H\'ebert et al.}
\graphicspath{{./}{figures/}}
\usepackage{macros}

\begin{document}

\title{Generation of realistic input parameters for simulating atmospheric point-spread functions at astronomical observatories}


\author{Claire-Alice H\'ebert \hskip2pt\href{http://orcid.org/0000-0002-7397-2690}{\includegraphics[width=9pt]{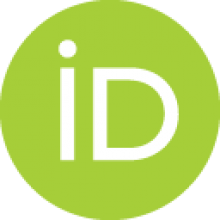}}}
\affiliation{Kavli Institute for Particle Astrophysics and Cosmology, Stanford University, Stanford, CA 94305}
\affiliation{Department of Applied Physics, 
Stanford University, Stanford, CA 94305}

\author{Joshua E. Meyers}
\affiliation{Kavli Institute for Particle Astrophysics and Cosmology, Stanford University, Stanford, CA 94305}
\affiliation{SLAC National Accelerator Laboratory, Menlo Park, CA 94025}

\author{My H. Do}
\affiliation{Department of Physics, Cal-State University Pomona, Pomona, CA 91768}

\author{Patricia R. Burchat}
\affiliation{Kavli Institute for Particle Astrophysics and Cosmology, Stanford University, Stanford, CA 94305}
\affiliation{Department of Physics, 
Stanford University, Stanford, CA 94305}

\author{The LSST Dark Energy Science Collaboration}

\begin{abstract}
High-fidelity simulated astronomical images are an important tool in developing and measuring the performance of image-processing algorithms, particularly for high precision measurements of cosmic shear -- correlated distortions of images of distant galaxies due to weak gravitational lensing caused by the large-scale mass distribution in the Universe.  
For unbiased measurements of cosmic shear, all other sources of correlated image distortions must be modeled or removed. 
One such source is the correlated blurring of images due to optical turbulence in the atmosphere, which dominates the point-spread function (PSF) for ground-based instruments.   
In this work, we leverage data from weather forecasting models to produce wind speeds and directions, and turbulence parameters, that are realistically correlated with altitude. 
To study the resulting correlations in the size and shape of the PSF, we generate simulated images of the PSF across a $\approx$ 10 square-degree field of view -- the size of the camera focal plane for the Vera C.~Rubin Observatory in Chile -- using weather data and historical seeing for a geographic location near the Observatory. 
We make quantitative predictions for two-point correlation functions (2PCF) that are used in analyses of cosmic shear.
We observe a strong anisotropy in the two-dimensional 2PCF, which is expected based on observations in real images, and study the dependence of the orientation of the anisotropy on dominant wind directions near the ground and at higher altitudes. 

The code repository for producing the correlated weather parameters for input to simulations (\psfws) is public at \url{https://github.com/LSSTDESC/psf-weather-station}.  
\end{abstract}

\section{Introduction} \label{sec:intro}

Images of distant galaxies are distorted by weak gravitational lensing due to inhomogeneities at large scales in the mass distribution in the Universe between the galaxy and the observer.  
These distortions are called cosmic shear \citep[see, \eg,][]{kilbinger_cosmology_2015, mandelbaum_weak_2018}. 
Spatial correlations in the distortions are sensitive to the properties and evolution of the matter density on large scales and the geometry of space.  
As the Dark Energy Survey \citep[DES\footnote{\url{https://www.darkenergysurvey.org/}},][]{amon_dark_2022, secco_dark_2022}, Hyper Suprime-Cam \citep[HSC\footnote{\url{https://hsc.mtk.nao.ac.jp/ssp/}},][]{li_hyper_2023, dalal_hyper_2023}, and Kilo-Degree Survey \citep[KiDS\footnote{\url{http://kids.strw.leidenuniv.nl/}},][]{asgari_kids-1000_2021} collaborations complete the analysis of their entire surveys, cosmic shear is becoming one of the most precise probes of cosmology -- in particular for measurements of the average matter density and the amplitude of matter fluctuations.
When the commissioning of the Vera C.~Rubin Observatory\footnote{\url{https://www.lsst.org/scientists}} is complete, the Legacy Survey of Space and Time (LSST) is expected to reach and then surpass the precision of existing surveys for cosmic shear by an order of magnitude. 

For accurate (unbiased) measurements of cosmic shear, all other sources of correlated image distortions must be modeled and/or removed. 
One such source is the correlated blurring of images due to Kolmogorov turbulence in the atmosphere, which dominates the point-spread function (PSF) for ground-based instruments such as DES, HSC, KiDS, and Rubin Observatory \citep[\eg,][]{heymans_impact_2012,jarvis_science_2016, xin_study_2018}. 
Algorithms are under active development for more accurately modeling and interpolating the PSF across the focal plane \citep[see, \eg,][]{bertin_automated_2011, jarvis_dark_2020}, correcting galaxy shapes for the PSF, and calibrating measures of cosmic shear from galaxy shapes \citep{sheldon_practical_2017, huff_metacalibration_2017, gatti_dark_2021, sheldon_metadetection_2023}.  
Both real astronomical images and high-fidelity simulated images are important tools for developing and optimizing these algorithms and measuring their performance, with simulations playing a unique role because the input parameters we are trying to measure or infer, including shear, are known.  

\begin{figure}
\includegraphics[width=0.45\textwidth]{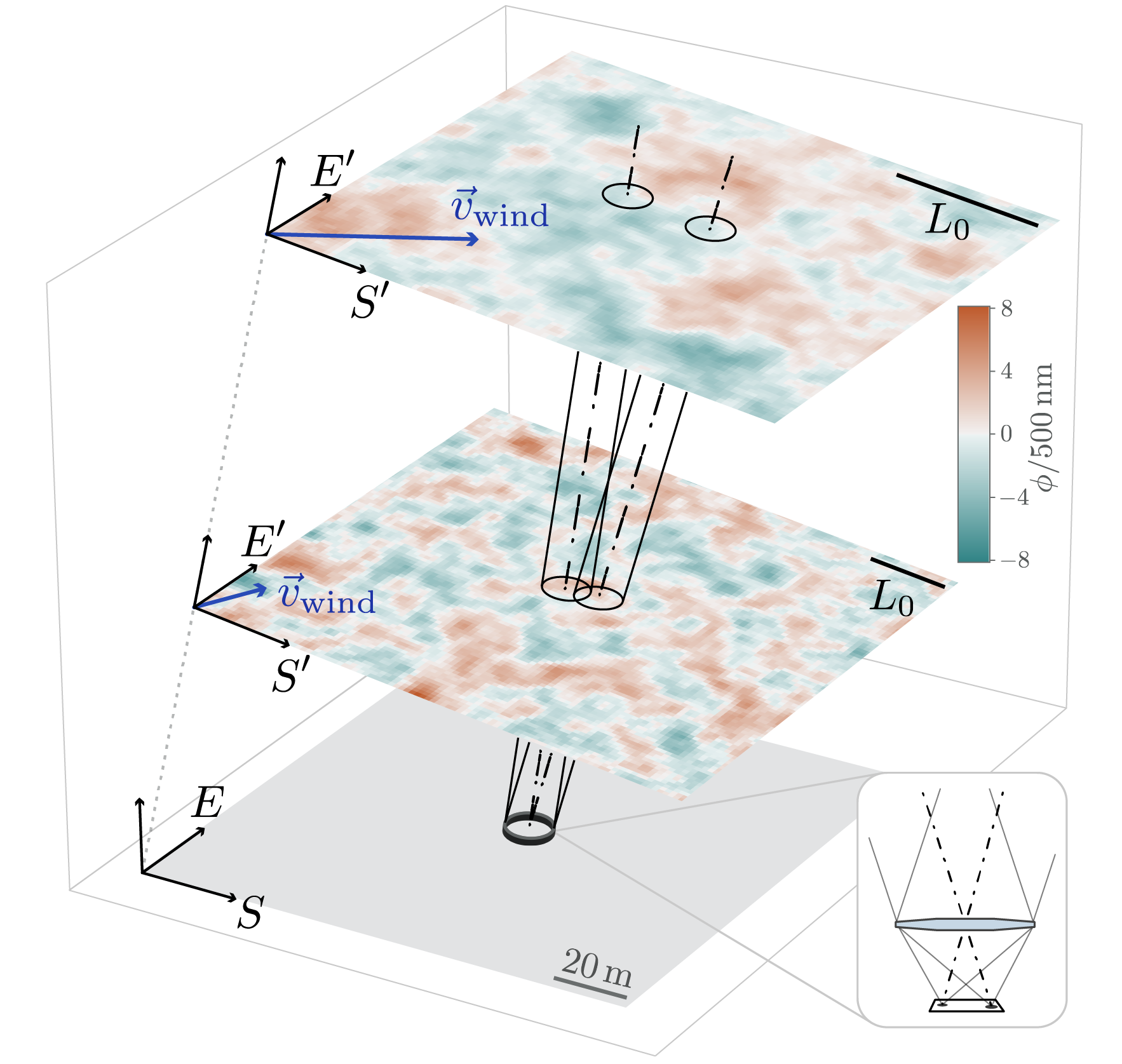}
\caption{
    This schematic illustrates the simplified view of the atmosphere used for PSF simulations based on discrete phase screens. 
    Lines of sight (dot-dashed black) for two stars (whose images are located at extrema of the field of view) pass through two phase screens of \vk refractive index variations, each with different values for the outer scale $L_0$. 
    The phase offset incurred by light passing through each point on each screen is indicated by the color scale in units of wavelength.
    The columns (solid black) associated with each line of sight show the path of starlight that will reach the telescope aperture (black), along with the relevant phase screen area. 
    The wind vectors (blue arrows) show speed and direction of the wind in the plane of the screen.
    The primed coordinate systems are perpendicular to the telescope axis, and are related to the ground coordinate system via the altitude and azimuth of the pointing.
    \label{fig:schematic}
    }
\end{figure}

High-fidelity atmospheric PSF simulations have played an important role in astronomy -- from the development of adaptive optics \citep{jolissaint_synthetic_2010, srinath_creating_2015, madurowicz_characterization_2018} to the optimization of instrumentation and software for precision cosmology, including Rubin Observatory and the associated survey, LSST  \citep{jee_toward_2011, chang_atmospheric_2012, peterson_simulation_2015, the_lsst_dark_energy_science_collaboration_lsst_2021}.
Simulations of the atmosphere often use a ``thin-screen, frozen-flow'' approximation in which the impact of a layer of atmosphere is modeled as a single two-dimensional planar ``screen'' in which the relative wavefront phase across the plane encodes the impact of variations in index of refraction in the layer of atmosphere (turbulence).  
The variation in phase across the screen is ``frozen'' during an exposure, but the screen moves across the field of view to simulate the impact of wind. 
The altitude dependence of wind speed and direction and of turbulence strength is modeled by including multiple screens, as illustrated in \figref{schematic}.

In preparation for analysis of data from Rubin Observatory, the LSST Dark Energy Science Collaboration (DESC) produced an extensive set of image simulations, called Data Challenge 2 \citep[DC2,][hereafter \dcii]{the_lsst_dark_energy_science_collaboration_lsst_2021}, in which turbulence parameters were based on median conditions determined in \cite{ellerbroek_efficient_2002} from measurements at \cp (the location of Rubin Observatory), and wind speeds and directions were drawn randomly from uniform distributions. 
In \cite{peterson_simulation_2015}, historical data are used for wind speeds and directions, and turbulence strengths; however, correlations between meteorological parameters at different altitudes, and the relationship between wind and turbulence parameters, are not taken into account.

As described above an important scientific goal in simulating atmospheric PSFs at Rubin Observatory or other wide-field instruments is to predict how weather in the observatory environment impacts correlations in PSF parameters across the focal plane. 
Therefore, it is important to include in the simulations altitude-dependent correlations among wind speed, wind direction, and turbulence, as well as realistic temporal variations.
We have produced a public software package (\psfws) that leverages local environmental telemetry and data products from global weather forecasting models to produce turbulence parameters and wind speeds and directions that are realistically correlated with altitude and in time. 
The package relies on an empirical model of atmospheric turbulence proposed in \cite{osborn_optical_2018} (hereafter \osborn), which parameterizes relative turbulence strength at different altitudes in the atmosphere as a function of the wind shear at that location. 

Whereas \osborn focuses mainly on fast predictions of turbulence as a function of altitude for real-time adaptive optics corrections in very large telescopes with narrow fields of view, our goal is to parameterize atmospheric conditions that, in an ensemble sense, are representative of a particular site and can be used to generate atmospheric PSFs across a wide field of view. 
We achieve this through the inclusion of weather-tower telemetry and site-specific empirical distributions of altitude-specific seeing contributions. 
We then use \psfws to predict and study the expected anisotropies in the PSF, the effects of different weather patterns, etc., at the location of a specific observatory.

In this paper, we present an application of the \psfws package to studies of the expected PSF at \cp (Rubin Observatory). 
However, the code is flexible enough to use for other observatories and includes functionality to download necessary datasets from weather forecasting services.

We describe how the atmosphere and PSFs are modeled in \secref{atmos}, outline the \psfws package in \secref{psfws}, describe three sets of inputs to simulations in \secref{imsiminputs}, define PSF parameters and two-point statistics in \secref{definitions2pcf}, and compare these PSF metrics for the three types of simulation inputs in \secref{imsimresults}and \secref{inputcompare}.
We end with a discussion of implications for cosmic shear analyses and future work. 

\section{Imaging through a turbulent atmosphere} \label{sec:atmos}
Since stars and galaxies are effectively light sources at infinity, their light can be treated as plane waves when entering the upper atmosphere. 
During the journey through the atmosphere, points on the surface of a wavefront accrue relative phase shifts.
The atmospheric component of an object's PSF is the result of the spatial variations in phase across the telescope pupil.

The phase shifts are caused by variations in the index of refraction in the atmosphere ($\delta n$) due to perturbations in air density driven by turbulent mixing of air at different temperatures  \citep{lawrence_survey_1970, clifford_classical_1978}.
These fluctuations in $n$ (referred to as \textit{optical turbulence}) vary in space and time; therefore, each photon in general incurs a slightly different cumulative phase shift on its path to the telescope pupil.
It is convenient to define an ``atmospheric column'', with diameter roughly that of the telescope pupil, which delineates the volume of turbulent air sampled by the imaged photons from a single source, as illustrated in \figref{schematic}. 
The atmospheric columns for each object in the field of view overlap at the pupil but diverge with distance from the telescope, resulting in a spatially varying, spatially correlated PSF over the focal plane. 

Optical turbulence exists for a range of spatial scales and amplitudes.
The spectral density of this turbulence, as a function of spatial frequency $\kappa$, can be described by the 3-dimensional \vk power spectrum \citep{von_karman_progress_1948, tokovinin_wavefront_1998}, where the subscript $n$ denotes index of refraction:
\begin{equation}\label{eqn:vk}
    E_n(\vect{\kappa};L_0) \propto ( \lvert \vect{\kappa} \rvert^2 + L_0^{-2})^{-11/6} \,.
\end{equation}
This is a modification of Kolmogorov's $\kappa^{-11/3}$ power law \citep{kolmogorov_local_1941}; the (altitude dependent) outer scale parameter $L_0$ sets an upper bound on the turbulence strength at low spatial frequencies, which would otherwise diverge as $\kappa \rightarrow 0$.
The \vk turbulence spectrum has an associated spatial correlation function (related to the Fourier transform of \eqnref{vk}). 
The degree of correlation for optical turbulence at a given altitude is set by the turbulence structure constant $C_n^2$; 
when given as a function of altitude $h$, this is known as the optical turbulence profile (OTP) $C_n^2(h)$. 
Although turbulent structures are constantly evolving with time, OTPs are typically assumed to be constant during the course of an exposure \citep{roddier_v_1981}.

Image quality is related to the OTP via the turbulence integral $J$:
\begin{equation} \label{eqn:j}
	J = \int_{h_1}^{h_2} C_n^2(h) dh \,.
\end{equation}
For the case when the integration bounds $h_1$ and $h_2$ correspond to the entire vertical extent of the atmosphere, $J$ quantifies the total strength of the turbulence experienced by photons passing through the corresponding atmospheric column.
$C_n^2$ and $J$ have units of $\unit{m^{-2/3}}$ and $\unit{m^{1/3}}$, respectively. 
The Fried parameter $r_0$ is a characteristic length scale that defines the radius of an aperture in which the wavefront phase variance is approximately $ 1\unit{rad^2}$ \citep{fried_statistics_1965}. 
It depends on $J$ as well as zenith angle $\zeta$ and wavenumber $k$ \citep{roddier_v_1981}:
\begin{equation} \label{eqn:r0}
	r_0 = (2.914 k^2 \sec \zeta J )^{-3/5}\,,
\end{equation}
and is inversely proportional to linear PSF size (FWHM).
In low turbulence conditions, $r_0$ is large (and $J$ is small), and thus the PSF size is small. 

The atmosphere is not a static system; in addition to turbulent mixing, there is large-scale motion driven by wind.
The component of wind velocity parallel to the telescope pupil  translates the optical turbulence across one or more atmospheric columns, leading to correlated phase shifts for photons with different positions and angles at the pupil plane, and therefore correlated PSF shapes across the focal plane.

\subsection{Atmospheric PSF simulations} \label{sec:sim}
Since it is computationally intractable to simulate atmospheric PSFs by calculating the trajectory of each photon through a turbulent (\ie, chaotic) 3D volume of atmosphere, approximate methods have been developed. 
In this section, we describe the frozen screen approximation, introduced in \secref{intro}, which has been used in the context of weak lensing studies; see \cite{jee_toward_2011, peterson_simulation_2015}; \dcii for more details.

Measurements of optical turbulence with SCIntillation Detection And Ranging (SCIDAR) instruments show that the atmosphere is often stratified into regions of stronger turbulence separated in altitude by areas of relative calm  \citep{osborn_optical_2018, osborn_atmospheric_2018}.
Typically only $\sim\,$1\unit{km} in vertical extent and variable in number, these layers of stronger turbulence dominate the atmospheric contribution to the PSF. 
These observations motivate a simplified model of the atmosphere that consists of only 2-dimensional phase screens across which the refractive index varies, with each screen representing a layer of turbulence.

The refractive index variations within each phase screen are a realization of \vk turbulence.
We assume Taylor's frozen flow hypothesis \citep{taylor_spectrum_1938}, in which the time scales for changes in turbulence are longer than those for changes due to phase screen drift from wind. 
Under this assumption, it is not necessary to evolve the turbulence structure during a simulated exposure. 
Instead, each phase screen is assigned a ``wind'' speed and direction; for each time step $\Delta t$ of the simulation, the phase screens are translated accordingly. 
A schematic of such an atmospheric simulation, with two phase screens, is depicted in \figref{schematic}.
After each time step, the phase screen contributions within the atmospheric column (for each star in the field) are summed vertically. 
These wavefront phase variations have a \vonkarman power spectrum (cf.\,\eqnref{vk} for $E_n(\vect{\kappa}; r_0)$): 
\begin{equation}
    E_W(\vect{\kappa};r_0,L_0) = 0.0228r_0^{-5/3}(|\vect{\kappa}|^2+\Lnot^{-2})^{-11/6}\,.
\end{equation}
The wavefront outer scale $\Lnot$ is the spatial scale at which correlations in wavefront phase saturate; it can be expressed as the turbulence-weighted sum of the turbulence outer scale $L_0$ over phase screens $i$: $\Lnot^{-1/3} = (\sum_i L_{0,i}^{-1/3}J_i)(\sum_iJ_i)^{-1}$ \citep{borgnino_estimation_1990, tokovinin_wavefront_1998}.
For each star, the wavefront is then Fourier transformed to focal plane coordinates and, after all time steps, added together to form the simulated image -- i.e., the PSF.
As a function of image coordinates $\theta_x,\theta_y$, 
\begin{equation}
    I(\theta_x,\theta_y) \propto \sum_{\Delta t} \left| \mathcal{F} \left\{ P(u,v) e^{-\frac{2\pi i}{\lambda}W(u,v, \Delta t)} \right\} \right|^2,
\end{equation}
where $P(u,v)$ is the aperture transfer function and $W(u,v,\Delta t)$ is the wavefront, with each a function of pupil coordinates $u,v$.
The sum over $\Delta t$ represents the sum over all simulation time steps during an exposure.

The phase screen is the building block of the simplified atmospheric model described above. 
Because turbulence integrals $J_i$ (\eqnref{j}) add linearly, each phase screen contributes to the total turbulence with weight $w_i = J_i(\sum_i J_i)^{-1}$.
Although one could use the turbulence integral $J_i$ to generate the phase pattern for screen $i$, it is more natural to use the Fried parameter $r_0$ because $r_0$ determines the turbulence power spectrum amplitude.
Given that $r_0 \propto J^{-3/5}$ (\eqnref{r0}),
the contribution of the $i$th screen is
$r_{0,i} =  w_i^{-3/5} r_0$.
By convention, $r_0$ is specified at $\lambda=500\unit{nm}$ and zenith angle $\zeta=0$. 

In summary, the input parameters for a simulation of PSFs across a single exposure are the outer scale $L_0$, the atmospheric seeing parameterized by $r_0$, the number of phase screens and their altitudes, the wind speed and direction for each screen, and the fractional contribution $w_i$ of each screen to the total turbulence. 

\section{\psfws} \label{sec:psfws}

In the \psfws software package, we leverage data products from weather forecasting organizations to produce realistically correlated wind and turbulence parameters. 
We use these input parameters to simulate correlated atmospheric PSFs across a large field of view, as described in later sections. 

\subsection{Motivation}
A potential source of bias in analyses of cosmic shear is uncorrected, spatially correlated noise \citep{mandelbaum_weak_2018}. 
The atmosphere is correlated via the \vk power spectrum described in \eqnref{vk} and, as we have seen, these spatial correlations translate into angular correlations in the size and shape of the atmospheric PSF in the associated exposure \citep{heymans_impact_2012, jarvis_dark_2020}.
Wind over the telescope plays an integral role in this process, as it moves correlated patches of turbulence through the atmospheric columns that  impact the images of different objects, leading to correlations on angular scales larger than the patches. 
If wind directions are consistent across altitudes, turbulence at different altitudes will imprint a stronger correlation in the PSF than when wind directions at different altitudes are uncorrelated. 

Another relevant factor for PSF correlations is the altitude dependence of the optical turbulence profile (OTP), which describes the contribution of each layer to the total turbulence strength \citep{roddier_v_1981}.
Interestingly, one of the drivers of atmospheric turbulence is wind itself -- specifically, wind shear \citep{masciadri_optical_2017}-- so we expect that these two factors that influence spatial correlations in PSF parameters are not independent.

\subsection{Data inputs to \psfws} \label{sec:inputs}

In \psfws, we separate the atmosphere into two regions based on the typical turbulence regime for those altitudes.
The ground layer (GL) is typically defined as the region between ground level and $500-1000\unit{m}$ above the telescope, where complex topography and heat sources generate non-Kolmogorov eddies.  
The free atmosphere (FA) is defined as the region above the ground layer, where turbulence is generally well-described by Kolmogorov statistics.
This separation into GL and FA plays an important role in many design choices for \psfws.

The primary sources of data for \psfws are data products from global weather forecasting organizations such as the European Centre for Medium-Range Weather Forecasts\footnote{\url{https://www.ecmwf.int/en/forecasts/datasets/}} (ECMWF) and the National Oceanic and Atmospheric Administration National Centers for Environmental Prediction\footnote{\url{https://www.emc.ncep.noaa.gov/emc/pages/numerical_forecast_systems/gfs.php}} (NOAA NCEP).
The global circulation models (GCM) used in weather forecasting cover the entire globe on a grid of 0.25-2.5\deg resolution and output predictions for dozens of environmental parameters at a number of different altitudes between 0 and $\approx$80\km above sea level.

Although any of the available GCM data sets can be used in \psfws, the results in this paper are based on data from the ECMWF Reanalysis v5 (ERA5) catalog; these are hourly estimates of global weather conditions based on meteorological measurements assimilated into the GCM.
ERA5 was chosen for its denser sampling both in time -- hourly -- which is useful for sampling conditions throughout the night, and in altitude -- 137 levels -- which is important for capturing vertical wind gradients in the atmosphere. 
We provide more details on the ECMWF and NOAA NCEP models in the Appendix.

The GCMs give us robust estimates of wind and temperature throughout the free atmosphere as a function of time. 
However, interactions of the atmosphere with the ground are not accurately captured because topographical features are not modeled at scales smaller than $\sim$1\km; 
therefore, the accuracy of the GCM data (both initial conditions and predictions) is limited near the ground.
We overcome this limitation in \psfws by using measurements from a weather tower on the telescope site, rather than GCM data, for the ground layer.
Since weather tower data are typically recorded every few minutes, the sampling times of the telemetry information can be matched to the GCM data used for the free atmosphere.
The weather tower measurements are optional inputs to \psfws but highly recommended since the GL is of significant importance for the PSF, contributing between 40 and 60\% of the turbulence at many observatories \citep{tokovinin_statistics_2003, tokovinin_model_2005, tokovinin_optical_2005}.

\subsection{Optical turbulence in \psfws}\label{sec:psfwsturb}
Some existing non-hydrostatic atmospheric models with sub-kilometer horizontal resolution are successful in simulating optical turbulence around observatories \citep{masciadri_3D_2001, masciadri_optical_2017}; however, such models are computationally prohibitive when many realizations need to be simulated. 
On the other hand, useful parameterizations of optical turbulence as a function of environmental quantities -- such as temperature, pressure, wind, and kinetic energy -- can be adapted from this literature, as is done in \osborn \citep{osborn_optical_2018}.
In particular, in \osborn,  wind shear is assumed to be the sole contributor to the kinetic energy term (\ie, wind shear is the source of turbulent mixing of air) and the temperature and pressure profiles from mesoscale simulations are replaced with GCM data.
In exchange for coarser resolution and more limited accuracy, with minimal computational time the \osborn empirical model produces estimates of $C_n^2(h)$ which, as shown in \osborn, are broadly consistent with stereo-SCIDAR measurements. 

The \osborn model captures variations in turbulence strength with altitude, but not the absolute strength; it requires calibration for the total turbulence $J$.
In addition, the \osborn model significantly under-predicts turbulence in the GL, which is expected since turbulence in the GL can have significant contributions from sources other than wind shear.
\psfws combines the \osborn optical turbulence model with complementary information from the literature to produce correlated turbulence parameters -- a value of $J$ for each phase screen, including the ground layer -- as described below.

Measurements of the altitude dependence of atmospheric turbulence with multi aperture scintillation sensor and differential image motion monitor (MASS-DIMM) instruments
at a variety of sites show that turbulence contributions from the FA and the GL are \textit{independent} \citep{tokovinin_model_2005, tokovinin_optical_2005, chun_mauna_2009}.
Motivated by this independence, the total turbulence in the GL and FA layers ($J_{\rm GL}$ and $J_{\rm FA}$) are treated separately in \psfws.
The relative amount of turbulence contributed by each FA layer is calculated with GCM data and \osborn, and the total GL and FA integrals are drawn from log-normal distributions fit to published quantiles of measurements of $J_{\rm GL}$ and $J_{\rm FA}$.

\begin{figure}
\includegraphics[width=0.47\textwidth]{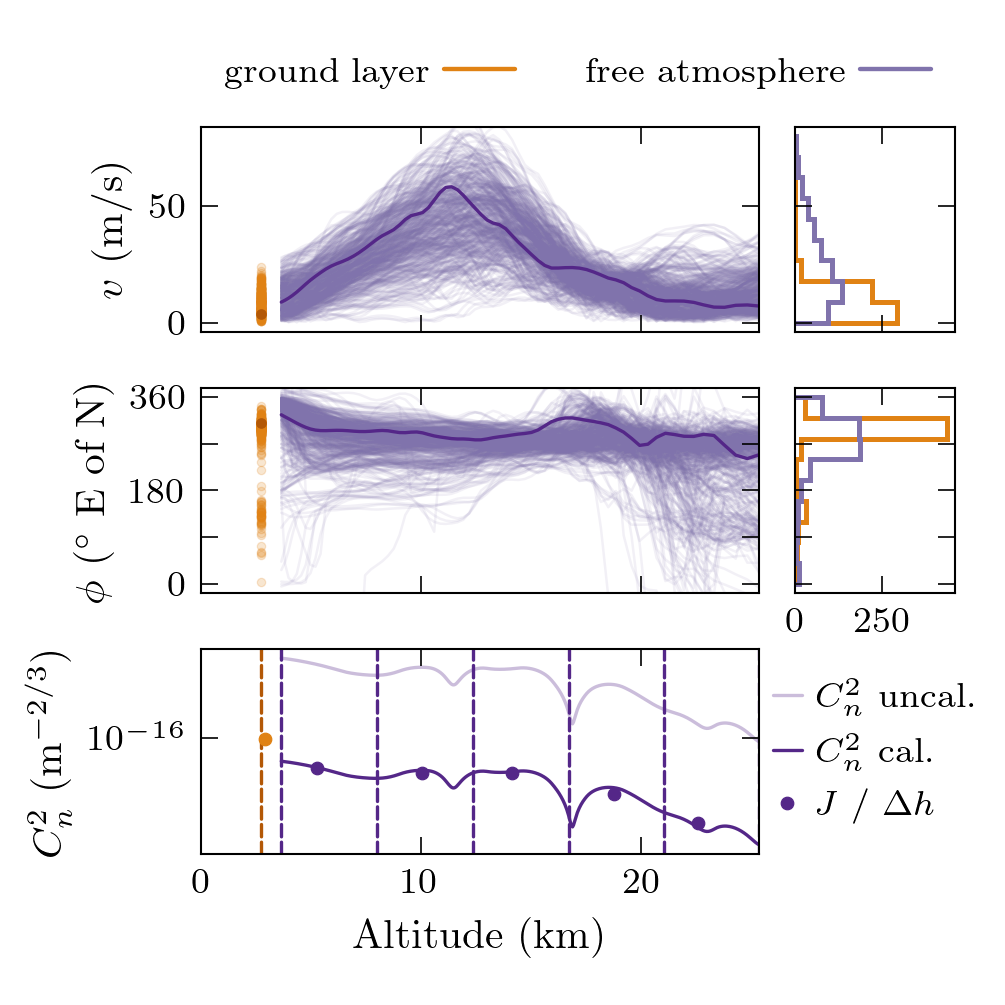}
\caption{
    Six months of wind data at and above \cp (May 2019 through  Oct 2019), processed with \psfws as described in \secref{psfws}.
    We plot wind speed (top) and meteorological direction (direction of wind \textsl{origin}; middle) as a function of altitude, and as a frequency distribution (right). 
    In the top two rows, weather tower measurements near the ground at Gemini South are shown in orange and ECMWF ERA5 data for the free atmosphere are shown in purple.
    The heavy purple line in each panel corresponds to data from a representative example time (August 6 2019, 08h UTC).
    In the bottom panel, the uncalibrated $C_n^2(h)$ profile for the same time is shown in light purple; the calibrated profile (scaled by $J_{\rm FA}$) is shown in dark purple. 
    The dashed vertical lines depict the boundaries between the altitude bins used to calculate turbulence integrals $J$ for each FA phase screen.
    The dots correspond to the value of $J$ divided by the corresponding range of altitudes and are placed at 200\unit{m} above the elevation of the observatory (orange dot) and the $C_n^2$-weighted position within each FA layer (purple dots).
    \label{fig:inputs}
    }
\end{figure}

In \figref{inputs}, we illustrate the steps taken to go from raw GCM data to simulation-ready input parameters in \psfws, with \cp as an example site.
Our data sources are ground layer telemetry from the weather tower at Gemini South (also located at \cp, about 1\km from Rubin Observatory), ECMWF ERA5 data products,\footnote{The closest grid point to \cp, at -30.241, -70.737, is $\approx 1.9\unit{km}$ away at -30.25, -70.75.} and $J_{\rm GL}$ and $J_{\rm FA}$ quantiles from \cite{tokovinin_model_2005}. 
The plots in the top two rows of the figure show six months of site telemetry (orange dots and histogram) and ERA5 data (purple curves and histogram). 
The GCM altitude profiles are sampled at 00h, 02h, and 04h local time; each profile has a corresponding co-temporal ground-layer data point.
The frequency distributions on the right illustrate the range of wind speeds and the distinctly non-uniform distribution of wind directions, both near the ground and across most altitudes.
In this paper and in \psfws, the direction is defined as the angle from which the wind is blowing, East of North. 

The heavy purple curve in each of the top two plots in \figref{inputs} corresponds to an example time that is representative of the median weather conditions. 
This representative example will serve to illustrate our process in this and the following sections.
The \osborn output for this representative example is shown in the bottom panel of \figref{inputs}; the light purple curve is the uncalibrated data and the dark curve is shifted to match a calibration value chosen randomly from the $J_{\rm FA}$ distribution.

Given the number $N$ of desired phase screens, \psfws places one in the ground layer, and the free atmosphere is divided into $N-1$ equally spaced layers.
Here, we choose to calculate turbulence integrals for $N=6$ layers of the atmosphere.
In the bottom panel of the figure, the purple vertical dashed lines correspond to the boundaries between the six layers; the orange dashed line corresponds to the altitude of ground, here 2737\unit{m} at Gemini-South.
The dots correspond to the values of the turbulence integrals ($J$) divided by the width of the layer, and are placed at the $C_n^2$-weighted position within each FA layer (purple dots) and at 200\unit{m} above the elevation of the observatory (orange dot) to match the location of the ground-layer screen in \dcii and thereby facilitate comparisons described in \secref{imsiminputs}.
The variable number of layers and the turbulence-weighted altitudes of the corresponding phase screens together offer some ability to model the temporal variability in the turbulence profiles observed in SCIDAR data (\osborn).
The complete set of parameters returned by \psfws is $h$, $J(h)$, $v(h)$, and $\phi(h)$, where $v(h)$ and $\phi(h)$ are the interpolated wind speed and direction at height $h$ of each  phase screen.
\psfws does not currently provide estimates for the outer scale $L_0$ due to the lack of physically motivated models linking environmental parameters to the outer scale.

Using random draws from turbulence distributions for ground turbulence integrals and calibration of FA $C_n^2$ is not an optimal solution since these draws are not temporally correlated with other environmental parameters. 
There is currently some observational evidence from a variety of observatories \citep{tokovinin_statistics_2003, tokovinin_optical_2005, chun_mauna_2009} for correlation existing between ground wind speed and $J_{\rm GL}$, so while we include in \psfws an option to correlate the random $J_{\rm GL}$ draws with ground wind speed, by default the correlation is set to zero.
There is also possible correlation of $J_{\rm GL}$ with ground wind \textit{direction} \citep{tokovinin_statistics_2003} at some observatories, but we have not yet implemented such an option in \psfws. 
Since there is only limited empirical evidence of correlations between $J_{\rm FA}$ and FA wind speeds \citep{tokovinin_statistics_2003} -- and in the \osborn model turbulence already depends on wind shear -- we do not include an option to introduce correlations. 

This method of using random draws from empirical distributions does somewhat restrict the predictive capabilities of simulations run with \psfws, as we do not expect to recover the average seeing on individual nights.
(Predicting the seeing on individual nights would require access to either MASS-DIMM or SCIDAR measurements, at or near the relevant observatory, that could be temporally matched to weather forecasting data products.)
However, we expect to recover overall seeing statistics as well as spatial correlations of the PSFs.

\section{Simulations of PSFs at Cerro Pach\'on}\label{sec:imsiminputs}
\psfws uses multiple sources of telemetry and vetted models to generate sets of correlated parameters for input to simulations of PSFs across the field of view.
In this section, we describe tests of these generated parameters, which aim to quantify how simulations that use as input \psfws parameters compare to earlier generations of atmospheric PSF simulations with uncorrelated parameters. 

All simulations described here are generated with the GalSim\footnote{\url{https://github.com/GalSim-developers/GalSim}} software library \citep{rowe_galsim_2015}.
We use the same GalSim implementation as described in \dcii, which follows ray-tracing methods developed in \cite{jee_toward_2011} and \cite{peterson_simulation_2015}.\footnote{The simulation includes, for each photon, a refractive kick proportional to local instantaneous phase-screen gradients to treat large scale turbulence, and a statistical ``second kick" treatment of small scale turbulence.  We do not enable any chromatic effects, the impact of the optical system is modeled as a simple Airy function, and the sensor is simulated as a perfect photon collecting grid.  No background light is included.}

We generate atmospheric PSF simulations for Rubin Observatory with three types of input parameters.

\begin{enumerate}
\item \psfwssims:
In the first case, input parameters are generated for \cp using \psfws with the data summarized in \figref{inputs}.
Six phase screens are used; the altitudes are allowed to vary according to the $C_n^2$ scheme described in \secref{psfwsturb}.

\item \bench:
As a second case, we use as a benchmark the input atmospheric parameters used in the DESC Data Challenge 2 (DC2) image simulations (see \dcii).
For each of six phase screens, the wind speed and direction are drawn from uniform distributions between 0-20\unit{m/s} and 0-360\degree, respectively.
Small ($\sim$10\%) Gaussian variations around the turbulence integrals from \cite{ellerbroek_efficient_2002} are introduced, but the associated six altitudes remain fixed between simulations.

\item \match: 
As a third case we use the same values of input parameters as in \bench\ -- \textit{except for the wind directions}, which are matched to the correlated wind directions used in \psfwssims.
The motivation for this \match simulation is to identify whether differences between distributions of PSF parameters for the first two cases are mainly driven by the highly correlated wind directions in \psfws.
\end{enumerate}

For each of the three cases (\psfwssims, \bench, and \match), we simulate one 30-second, 3.5-deg exposure (the expected exposure time and field of view for Rubin LSST) for each of the 531 time points in the six months of ERA5 and site telemetry data. 
Each triplet of simulations has the same outer scale (drawn from a truncated log normal distribution with median of $25\unit{m}$ and used for all phase screens), the same random seed for realizing the turbulent phase variations in the phase screens, and the same atmospheric seeing (drawn uniformly from 0.6 to 1.6\asec). 
The contribution of seeing from each phase screen varies according to the turbulence integrals used for that case (\psfws outputs for \psfwssims; randomized \cite{ellerbroek_efficient_2002} for \bench and \match).

PSF images are generated at 50k random locations across the field of view with a pixel resolution of 0.2 \asec. 
Each PSF is drawn with sufficient photons ($10^6$) such that Poisson fluctuations are not significant and then convolved with a Gaussian of 0.35\asec FWHM to account for the PSF contribution from optics and sensors.
To avoid issues related to overlapping PSF images, each PSF is generated and measured individually on a $50\times50$ pixel grid.

\section{PSF Parameters and two-point statistics}\label{sec:definitions2pcf}

We estimate PSF size and shape from the (weighted) second moments $Q_{ij}$ of each PSF intensity profile $I(\theta_x, \theta_y)$:
\begin{equation}
    Q_{ij} = \frac{\int d^2\theta \, I(\theta_x,\theta_y) \, W(\theta_x,\theta_y) \, \theta_x \, \theta_y }{ \int d^2\theta \, I(\theta_x,\theta_y) \, W(\theta_x,\theta_y)},
\end{equation}
where $\theta_x$ and  $\theta_y$ correspond to angular position on the focal plane and $W(\theta_x,\theta_y)$ is a weighting function. 
We have used the \galsim implementation of the HSM adaptive moments algorithm \citep{hirata_shear_2003} to measure PSF $Q_{ij}$.

As a measure of PSF size, we use 
\begin{equation}
    \sigma = {\rm det}(Q)^{1/4}. 
\end{equation}
For PSF shape, we use a definition of ellipticity $e$ that is commonly used in weak lensing analyses\footnote{See Part 3, Fig. 2 in \cite{schneider_gravitational_2006}, for example.}: 
\begin{equation}
    e = e_1 + i e_2 = \frac{Q_{xx} - Q_{yy} + 2iQ_{xy}}{Q_{xx} + Q_{yy}}. 
    \label{eqn:e_def12}
\end{equation}
The magnitude of $e$ is given by  $|e| = \frac{1-q^2}{1+q^2}$, where $q$ is the ratio of the minor to major axis of the second-moment ellipse. 
If $e_2=0$, then the orientation of the major axis of the ellipse is parallel to or perpendicular to the $\theta_x$ direction for positive or negative values of $e_1$, respectively. 
If $e_1=0$, then the major axis of the ellipse lies parallel to or perpendicular to an axis rotated $+45^\circ$ with respect to the $\theta_x$ axis, for positive or negative values of $e_2$, respectively.
Non-zero $e_1$ and $e_2$ describe orientations in between. 

This (complex) ellipticity parameter $e$ has a well-defined response to lensing shear when averaged across an ensemble of galaxy images if the effect of the PSF on the image has been accurately removed. 
Errors in the model for the PSF size and shape result in multiplicative and additive shear biases, respectively, and the exact impact on the ensemble weak lensing shear observables also depends on their spatial correlations.\footnote{For a review of weak lensing requirements for precision cosmology, see \cite{mandelbaum_weak_2018}.} 

Spatial two-point correlation functions (2PCFs) for both PSF size and shape are relevant in weak lensing analyses. 
We define the PSF size two-point correlation function as 
\begin{equation}
    C(\theta) = \langle \delta\sigma_a \, \delta\sigma_b\rangle(\theta),
    \label{eqn:shape_corr}
\end{equation}
where $\delta\sigma$ is the deviation in PSF size $\sigma$ from the mean size in that exposure, the indices $a$ and $b$ denote a pair of PSFs, and the angle brackets indicate an average over all pairs at each angular separation $\theta$ in the field of view. 

For calculations of two-point correlation functions for PSF shape parameters, it is most useful to define the complex ellipticity with respect to the separation vector between each pair of PSFs:
\begin{equation}
e = e_t + i e_\times,
\label{eqn:e_deftx}
\end{equation}
where the tangential  and cross components $e_t$ and $e_\times$ play the roles of $e_1$ and $e_2$, respectively, in \eqnref{e_def12}. 
In other words, the $y$ axis in \eqnref{e_def12} is now oriented along the separation vector and the $x$ axis perpendicular to it.

A pair of two-point correlation functions for PSF shape are defined in terms of the tangential and cross components of $e$: 
\begin{equation}
    \xi_\pm(\theta) = \langle e_{t,a} e_{t,b} \pm e_{\times,a} e_{\times,b} \rangle(\theta). 
\end{equation}

\begin{figure}
\includegraphics[width=0.47\textwidth]{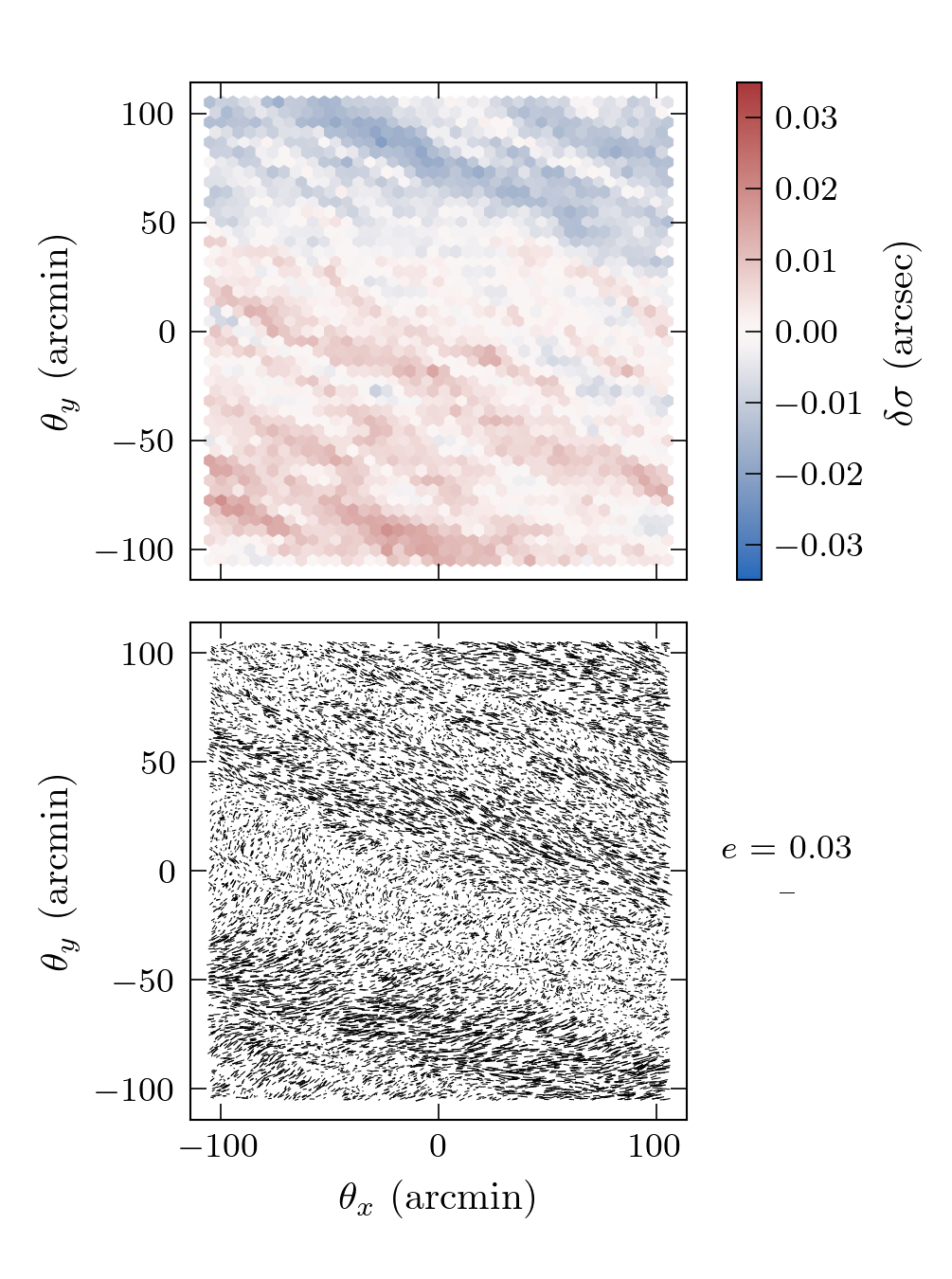}
    \caption{
    Spatial distributions of PSF size relative to mean size (top) and PSF ellipticity (bottom) across a simulated 3.5-\unit{deg^2} exposure generated with the \psfws input parameters for the representative example depicted by the dark curves and dots in \figref{inputs}.   
    The orientation of each line in the bottom plot corresponds to the orientation of the major axis of the PSF shape, while the length and color contrast are proportional to the magnitude of the ellipticity $e$ defined in \eqnref{e_def12}. 
    \label{fig:output}
    }
\end{figure}

\begin{figure}
\includegraphics[width=0.47\textwidth]{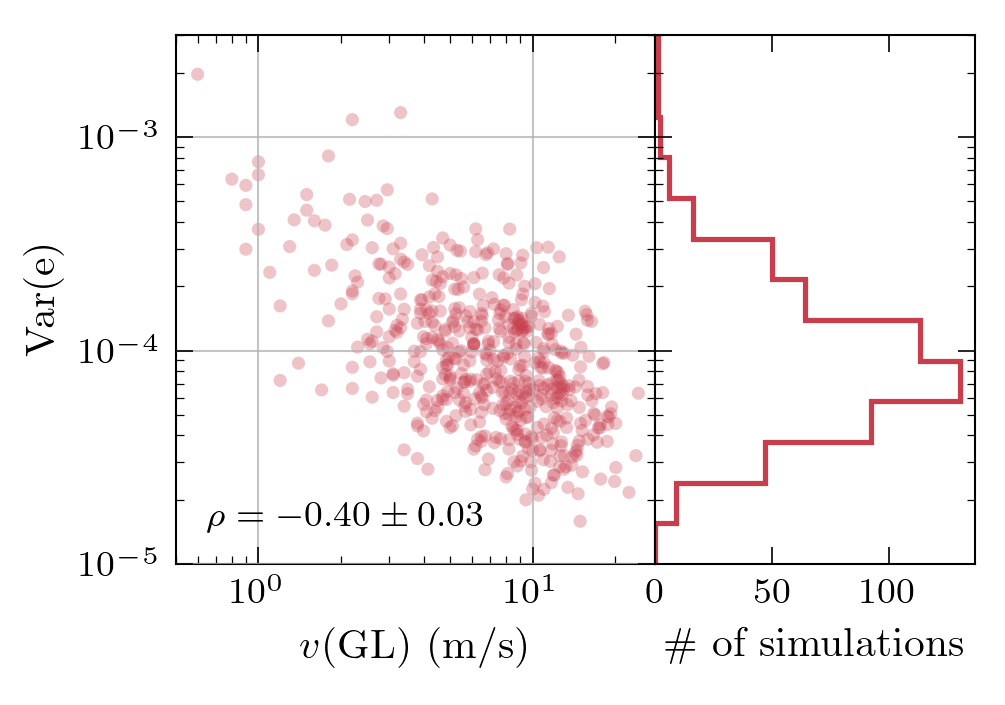}
    \caption{Variance of PSF ellipticity $e$ for each simulated exposure versus the ground layer (GL) wind speed in that simulation, for the \psfwssims simulations described in \secref{imsiminputs}. 
    The Pearson correlation coefficient $\rho$ is reported on the scatter plot.
    \label{fig:paramvar}
    }
\end{figure}

\section{Simulation results and comparisons} \label{sec:imsimresults}

For each type of simulation described in \secref{imsiminputs} -- \psfwssims, \match, or \bench\ -- we expect the simulated turbulence to imprint structure in the distribution of PSF parameters across a given exposure. 
As an example of the output of the \psfwssims simulation, we display in \figref{output} the spatial distribution of the PSF size relative to the mean size (top panel) and the PSF shape ($|e|$ and orientation, bottom panel) of the simulated PSFs for a single exposure generated with one of the \psfwssims parameter sets (the representative example described in \secref{psfwsturb} and depicted by the dark curves and dots in \figref{inputs}). 
The PSF parameters are clearly correlated across the field of view and the correlations are not isotropic.

In this section, we quantify these correlations and their anisotropy for each type of simulation inputs. 
We compare ensemble statistics across exposures for each simulation type, with a focus on the spatial two-point correlations of PSF size and shape, and study the dependence on particular input weather parameters.

\subsection{Variance in PSF parameters}
\label{sec:variances}

In \figref{paramvar}, we plot the variance of PSF ellipticity $e$ (defined in \eqnref{e_def12}) across each \psfwssims\footnote{The \bench and \match results (not shown) are consistent with the \psfwssims behavior described here.} simulated exposure versus the input ground-layer wind speed for that exposure.
The variance decreases with increasing ground-layer wind speed $v({\rm GL})$; 
a similar negative correlation with $v({\rm GL})$ is observed for variance in PSF size.  
This is expected because wind moves the  phase screens across the field of view, washing out variations in PSF size or shape due to turbulence structure at all angular scales; therefore, the higher the wind speed, the more the variations are suppressed. 

Correlations between variance in PSF size or shape parameters and free-atmosphere wind speed (not shown) are, in general, found to be weaker. 
This is expected because, instantaneously, different PSFs across the focal plane are more likely to sample independent regions of the phase screen with increasing altitude (see illustration in \figref{schematic}); 
the resulting variance in a PSF parameter is less likely to decrease due to motion of the screen, compared to the GL screen where different PSFs are more likely to sample overlapping regions of the phase screen.  

\subsection{1D two-point correlation functions}
\label{sec:covariances}
\begin{figure}
\includegraphics[width=0.47\textwidth]{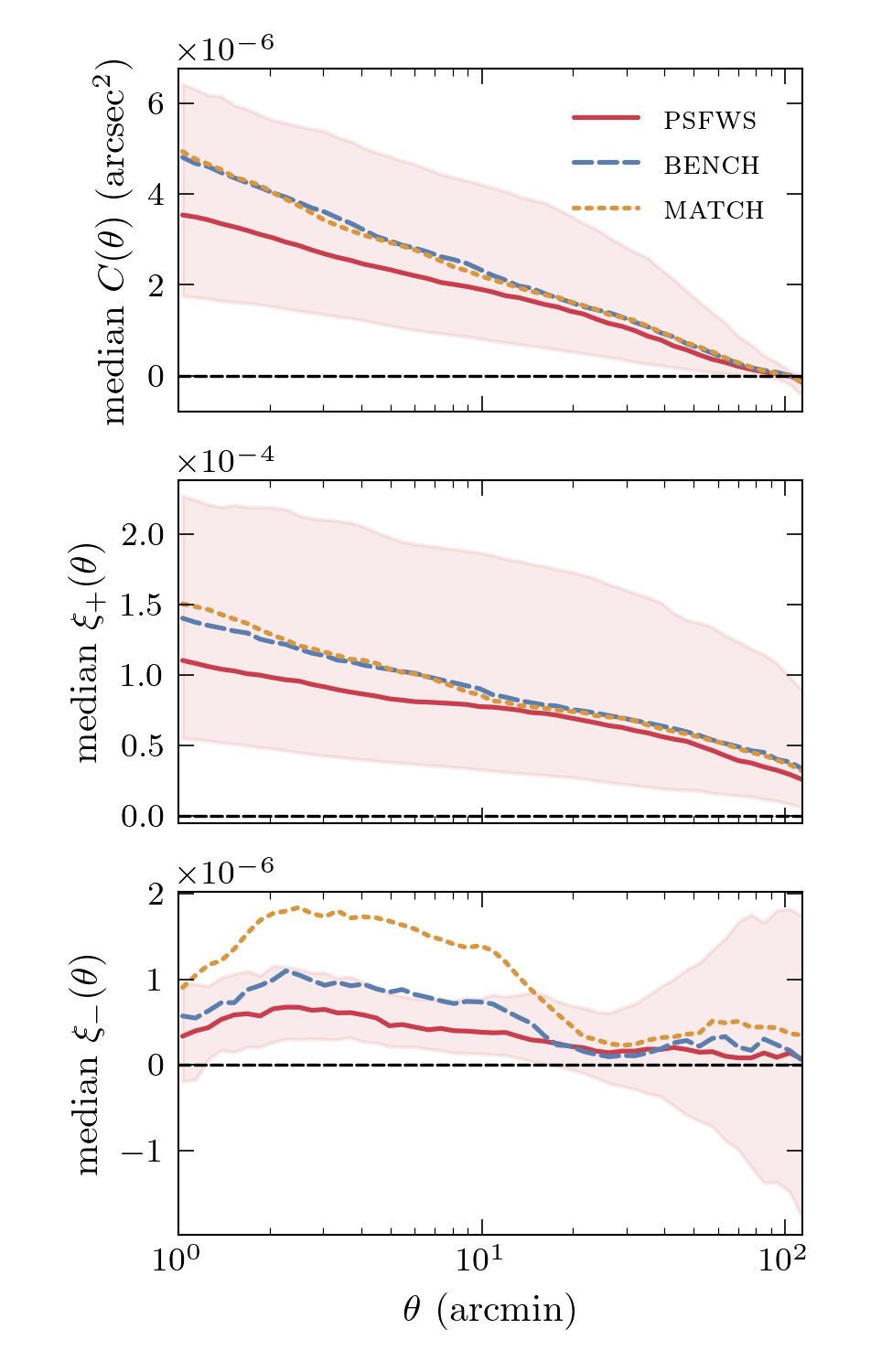}
    \caption{
    Values of the two-point correlation functions for PSF size ($C(\theta)$, top) and PSF ellipticity ($\xi_+(\theta)$, middle, and $\xi_-(\theta)$, bottom), as a function of angular separation $\theta$ between pairs of PSFs. 
    Ensemble median values are shown as curves for three different sets of inputs to the simulations, described in \secref{imsiminputs}.  
    The \psfws input parameters (wind speed, wind direction, and turbulence contribution) are correlated between phase screens at different altitudes (\psfwssims, solid red);
    the benchmark input parameters are not correlated between phase screens (\bench, long-dashed blue); 
    and the matched input parameters are the same as \bench but with phase-screen wind directions matched to those in \psfws (\match, short-dashed yellow).  
    Shaded areas depict region between $25$th and $75$th percentile values for \psfwssims simulation inputs (\ie, central $50$ percentile values).
    \label{fig:1d2pcf}
    }
\end{figure}

In \figref{1d2pcf}, we show the value of the two-point correlation function (2PCF) for PSF size ($C(\theta)$, top) and PSF ellipticity ($\xi_+(\theta)$, middle; $\xi_-(\theta)$, bottom), as a function of angular separation $\theta$ between pairs of PSFs.\footnote{All 2PCFs were computed using the TreeCorr software \citep{jarvis_skewness_2004}: \url{https://rmjarvis.github.io/TreeCorr}}  
The ensemble median values of the 2PCFs are shown as curves for each type of simulation input: \psfwssims (solid red), \bench (long-dashed blue), and \match (short-dashed yellow).
The shaded areas depict the region between the $25$th and $75$th percentile values of the 2PCFs for \psfwssims simulation inputs (\ie, the central $50$ percentile values).

The range of $\theta$ ($\sim$\,1 to 100\amin) is limited near the low end by the density of PSFs ($10^4$ per $4.4\times 10^4\amin^2$) and near the upper end by the size of the field of view ($\approx$\,210\amin ). 
The range of angular separations of interest in cosmic shear analyses is also $\sim$ 1 to 100\amin \citep[see, for example,][]{asgari_kids-1000_2021,amon_dark_2022, li_hyper_2023}.

The values of $\xi_+(\theta)$ at \smallsep are of order $10^{-4}$, which is of the same order (or larger than) the values expected and measured for cosmic shear.  
Since most galaxies used to probe cosmic shear are of similar size to the PSF, this implies that PSF shapes and their variation across the focal plane must be modeled accurately to avoid significant bias on measures of cosmic shear.  
Errors in the modeling of PSF size and its variation across the focal plane can also lead to biased shear 2PCFs \citep{rowe_improving_2010,jarvis_science_2016}.

For all three types of inputs to the simulations (\psfwssims, \bench, and \match), the values of $\xi_-(\theta)$ are approximately two orders of magnitude lower than those for $\xi_+(\theta)$), with median values that are slightly positive.

\subsection{2D two-point correlation functions}
\label{sec:anisocovar}
\begin{figure}
\includegraphics[width=0.5\textwidth]{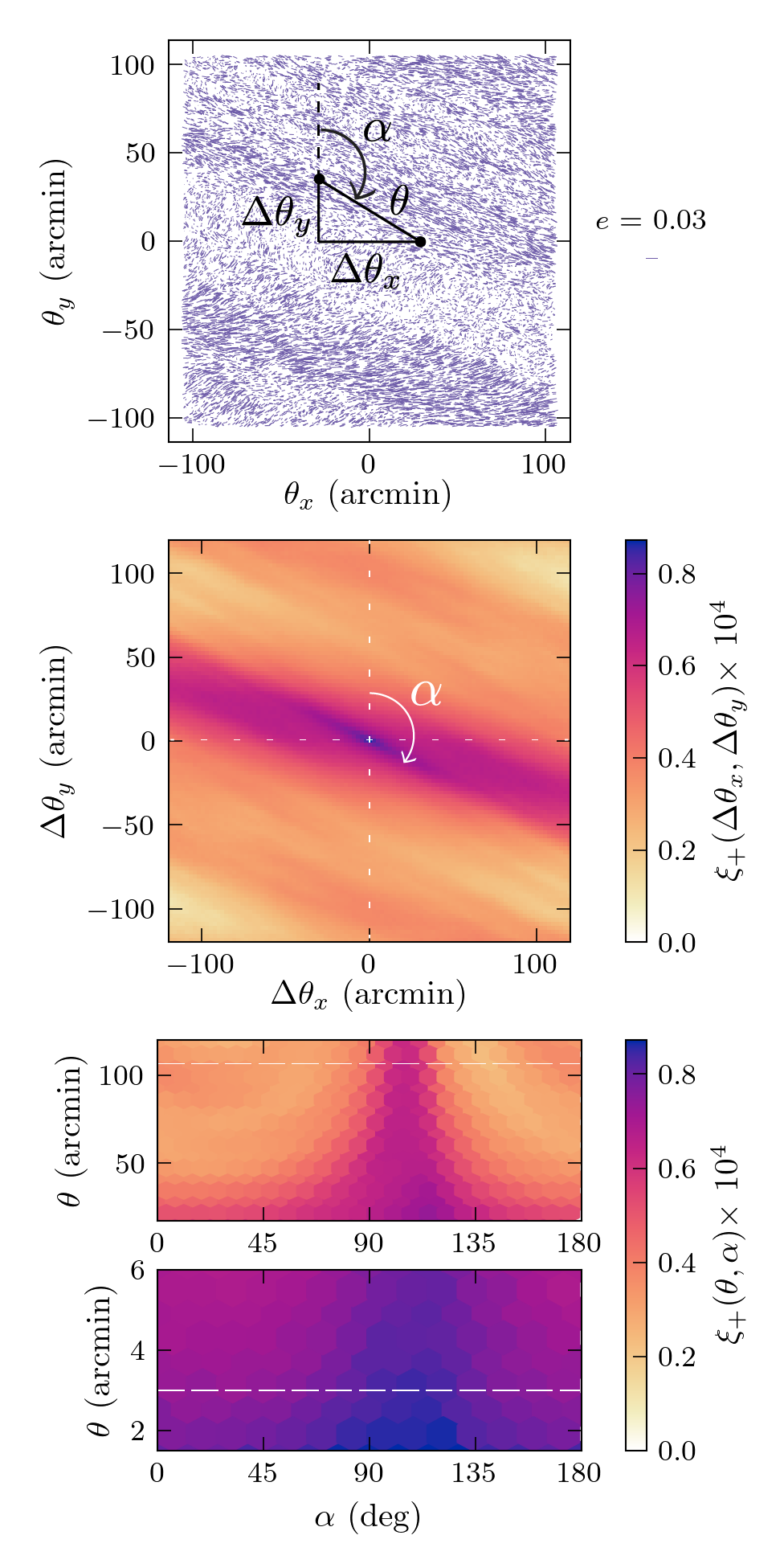}
    \caption{Illustration of anisotropic two-point correlation function of PSF shape for the \psfwssims representative example. 
  The top panel shows the simulated PSF ellipticity across the field of view overlain with a diagram of an example pair of PSF locations (black dots) and the separation between them, in Cartesian and polar coordinates.
  The color scale in the second panel shows the dependence of the 2PCF $\xi_+$ on the coordinates $\Delta \theta_x$ and $\Delta \theta_y$.
  The color scales in lower two panels show the same 2PCF transformed to polar coordinates (separation $\theta$ versus angle $\alpha$)  for two ranges of $\theta$: 20 to 120\amin, and 1.5 to 6\amin. 
  The range above the white dashed line in each of the two bottom panels is referenced in \secref{inputcompare}.
    \label{fig:anisoexample}
    }
\end{figure}

In order to describe and quantify anisotropies in the distribution of PSF parameters across the focal plane (such as those that are evident for PSF size and ellipticity in \figref{output}, and repeated for ellipticity in the top panel in \figref{anisoexample}), we introduce the angle $\alpha$ to describe the polar angle of the separation vector $\vec\theta$ between the location of two PSFs, measured with respect to the $\theta_y$ axis, as illustrated in the top panel in \figref{anisoexample}.
We will specify $\alpha$ in degrees (from 0 to $180^\circ$) and  will continue to specify angular separation $\theta$ in\amin. 

Since the magnitude of the 2PCF $\xi_+$ is approximately two orders of magnitude greater than that for $\xi_-$, we focus on quantifying anisotropies in $\xi_+$.
In the lower three panels in \figref{anisoexample}, the color scale is constant and depicts the value of $\xi_+$ for the single exposure simulated with the representative example of \psfws inputs used to generate the top panel.
The first of the lower three panels shows the dependence of $\xi_+$ on $\Delta\theta_x$, $\Delta\theta_y$.\footnote{The values of $\xi_+(\Delta\theta_x, \Delta\theta_y)$ are symmetric under $(\Delta\theta_x, \Delta\theta_y) \rightarrow (-\Delta\theta_x, -\Delta\theta_y)$; hence, $\alpha$ is defined in the range 0 to $180^\circ$.}   
On all scales, $\xi_+(\Delta\theta_x, \Delta\theta_y)$ is largest for pairs of PSFs  with a separation vector oriented at an angle $\alpha \approx 110^\circ$. 
In addition, for small separations ($\theta \sim$ 1\amin), $\xi_+$ is enhanced for pair orientations with $\alpha \approx 75^\circ$. 

In the lower two panels in \figref{anisoexample}, we show the same 2PCF but now in ``polar'' coordinates -- i.e., as a function of PSF pair separation $\theta$ and orientation angle $\alpha$ --  for two ranges of $\theta$: 20 to 120\amin, and 1.5 to 6\amin. 
In each plot we see a dark vertical band corresponding to the maximum values of \xip at orientations consistent with the above estimates of $\alpha\approx 110^\circ$ and $ \approx 115^\circ$ for large and small scales, respectively. 
For the same exposure, similar features are observed at the same angles $\alpha$ in plots of the 2PCF for PSF size, $C(\theta,\alpha)$, for large and small separations.

\section{Dependence of anisotropies on input wind parameters} \label{sec:inputcompare}

In this section, we probe the relationship between simulation input wind directions and the directions $\alpha$ along which the anisotropic 2PCFs for the output PSF parameters are maximum.
To quantify the orientation of the anisotropy in $\xi_+(\theta)$ at large and small separations, we identify the value of $\alpha$ for which \xip is maximum (denoted by $\alpha_{\rm max}$) for two ranges of separation: $\theta$ between 108 and 120\amin and between 3 and 6\amin, approximately coinciding with the largest and smallest scales currently used in cosmological analyses \citep[\eg, ][]{amon_dark_2022}. 
These ranges correspond to the region above the white dashed line in each of the lower two panels in \figref{anisoexample}. 
We find $\alpha_{\rm max}$ for each separation range for each of the 531 simulated exposures for each of the three types of simulation inputs (\psfwssims, \bench, \match) described in \secref{imsiminputs}. 

For wind direction, we use both the orientation $\phi(\text{GL})$ of the ground-layer wind velocity and the orientation $\phi(\text{FA})$ of the wind velocity in the free-atmosphere layer with highest turbulence strength.

\begin{figure*}
\includegraphics[width=0.99\textwidth]{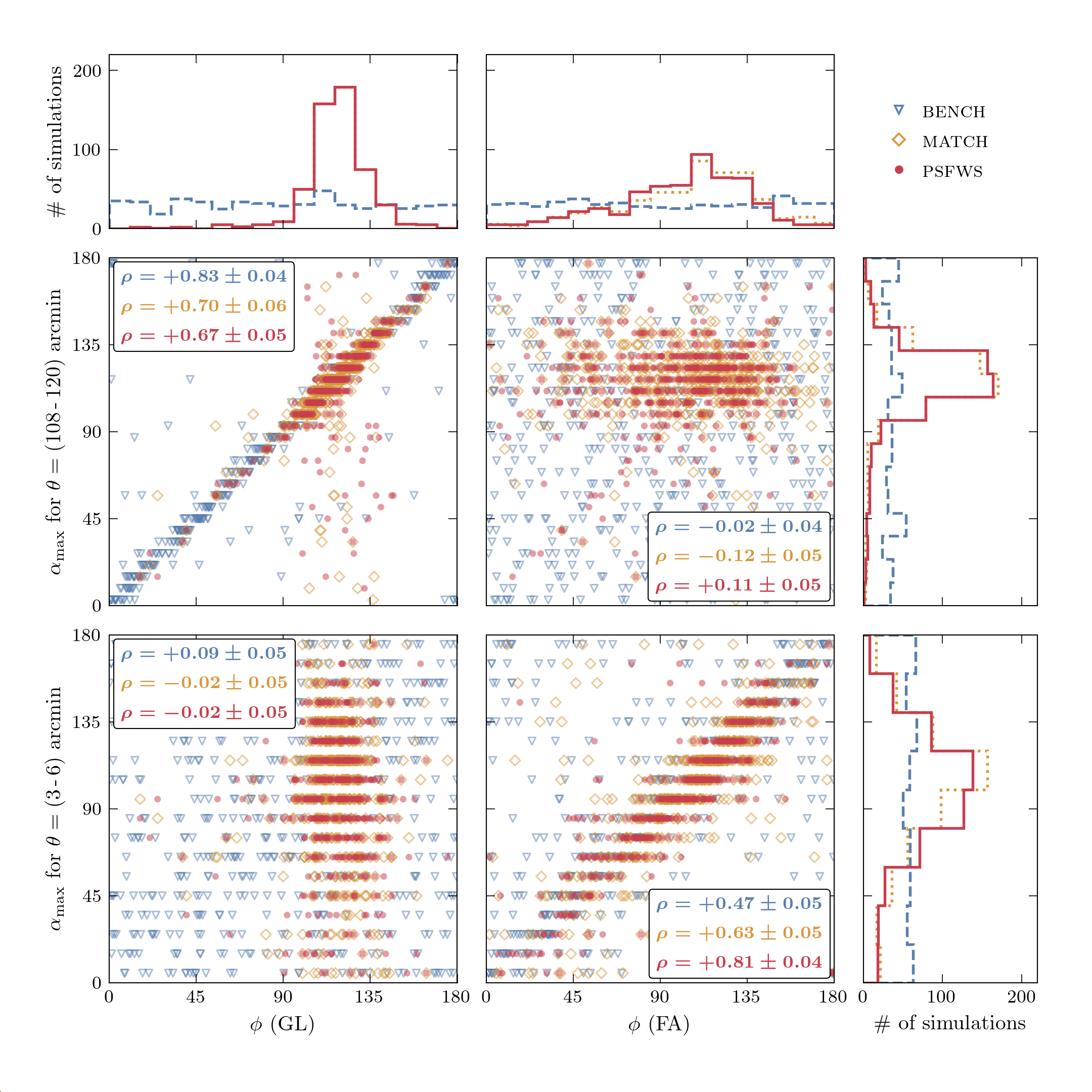}
    \caption{Orientation angle $\alpha_{\rm max}$ at which the anisotropic two-point correlation function $\xi_{+}$ for the PSF shape is maximum versus input wind direction $\phi$ for each simulated exposure, for the three types of simulation inputs (\bench, \match, and \psfwssims).
    The plots in the middle row correspond to large angular separations ($\theta$ between 108 and 120\amin) and those in the bottom row to small angular separations ($\theta$ between 3 and 6\amin).
    The angles $\phi({\rm GL})$ and $\phi({\rm FA})$ correspond to the wind velocity direction in the ground layer (left column) and in the highest-turbulence-strength free-atmosphere layer (middle column), respectively, modulo $180^\circ$. 
    The distributions of the angles are shown in the four projected histograms.
    Correlation coefficients for each type of simulation are reported on each scatter plot (\textsc{bench} in blue, \textsc{match} in gold, \textsc{psfws} in red).
    \label{fig:bigscatter}
    }
\end{figure*}

In the scatter plots in \figref{bigscatter}, we plot $\alpha_{\rm max}$ for the shape 2PCF $\xi_+(\alpha)$ versus $\phi(\text{GL})$ (left) and versus $\phi(\text{FA})$ (right), for separations of $\sim\,$100\amin (top) and $\sim\,$1\amin (bottom).  
The projected histograms in the top row illustrate how the distributions of GL and FA wind orientations are uniform for \bench simulations and peaked at the same dominant wind orientations for \psfwssims and \match simulations.\footnote{The distributions of $\phi(\text{GL})$ are identical for the \psfwssims and \match (matched) simulations by definition of the \match simulations. The distributions of $\phi(\text{FA})$ are slightly different for the \psfwssims and \match simulations because the turbulence profiles are not matched.}  

The Pearson correlation coefficient for each set of simulations is displayed on each scatter plot. 
A strong correlation between $\alpha_{\rm max}$ for \largesep and $\phi(\text{GL})$ (top left scatter plot) exists for all three types of simulation inputs. 
In contrast, for \smallsep, there is no significant evidence for correlation between $\alpha_{\rm max}$ and $\phi(\text{GL})$.

The results are quite different for the FA wind direction; there are no significant correlations at separations of $\sim\,$100\amin, but correlations exist at separations of $\sim\,$1\amin with some variation between the three types of inputs.
In particular, the correlation is very strong for simulations with realistic altitude-dependent FA wind directions and turbulence strengths (\psfwssims, $\rho=0.81\pm0.04$) but is lower when these parameters are chosen randomly with altitude (\bench, $\rho=0.47\pm0.05$). 
When input FA wind directions (but not turbulence strength) are matched to those of \psfwssims, the correlation coefficient has an intermediate value (\match, $\rho=0.63\pm0.05$). 
Although the correlation coefficients are statistically quite significant at \smallsep when FA wind directions are correlated across altitudes, the distributions of $\alpha_{\rm max}$ have higher variance than those at separations of $\sim100$\amin.  

To understand the FA results, we describe the expected angular scale of PSF correlations and a physical source of anisotropies. 

{\bf Angular scales:}
Atmospheric features of a particular \textit{physical} size that are located at lower altitudes have a larger \textit{angular} size in the focal plane than features of the same physical size at higher altitudes. 
Therefore, at one instant in time an optical turbulence pattern near the ground results in spatial variations in the PSF over larger angular separations than the same turbulence pattern located near the top of the free atmosphere. 
Consider, for example,  turbulence with a scale equal to the median outer scale $L_0$. 
(The power spectrum flattens for spatial frequencies lower than $\sim 1/L_0$, as shown in \eqnref{vk}.) 
For $L_0 = 25\unit{m}$ and a ground-layer screen at a height of  200\unit{m}, the angular scale of turbulence variations is $\arctan\frac{25}{200}\approx 400\amin$; 
for a phase screen at an altitude of 20\unit{km}, the angular scale is $\arctan\frac{25}{20000}\approx 4\amin$.
This illustrates why the ground layer is more relevant at large angular separations and the free atmosphere is more relevant as small separations.

{\bf Sources of PSF anisotropies and correlations with wind:}
While wind speeds and directions can change throughout the night, they are typically fairly stable during the course of an exposure. 
This is particularly true at \cp, where the wind direction is persistently coming from the sea.
During a 30-sec exposure then, we expect air at all altitudes to be moving coherently; different layers may be translating at different speeds, but in mostly the same direction.
As a consequence, PSF images along the direction of the wind on the focal plane have ``seen'' much of the same optical turbulence, although in slightly different combinations due to wind speed variation with altitude. 
The shapes of these PSFs will thus be more correlated with each other than with those in a direction orthogonal to the wind.

The realistic inputs used to generate the \psfwssims exposures include these  correlated wind speeds and directions. 
In the case of \bench simulations, there is no coherent motion of the turbulence in the FA because the wind speed and wind direction for each layer is chosen randomly; therefore, the direction of the highest-turbulence layer is not related to that of the other FA layers.
This results in a correlation coefficient $\rho$ between FA wind direction and $\alpha_{\rm max}$ that is suppressed relative to \psfwssims at separations of $\sim\,$1\amin.
The \match simulations have wind directions that are correlated across altitudes, but speeds are chosen randomly. 
The random speeds cause turbulence layers to move with greater differences in speed than the smoothly varying wind profiles in \psfwssims, resulting in a slightly suppressed correlation coefficient for \match simulations despite the wind directions being matched with \psfwssims.

These conclusions hold when considering the ensemble of FA layers: we found similar results, albeit with less correlation, when using the sum of the turbulence-weighted velocities of all FA layers to define $\phi({\rm FA})$, rather than choosing the velocity of the single layer with highest turbulence.

Similar correlations are also observed between wind direction and the orientation of anisotropies for PSF size 2PCF $C(\alpha)$, although the values of the correlation coefficients between $\alpha_{\rm max}$ and $\phi(\text{GL})$ at separations of $\sim\,$100\amin are lower than observed for PSF shape, potentially due to noisier estimates of $\alpha_{\rm max}$.

\section{Conclusions and Future Work} 
As described in the introduction, accurate measures of cosmic shear with future astronomical surveys, such as LSST at Rubin Observatory, require unbiased measures of two-point statistics for galaxy shapes, which in turn require unbiased measures of the size and shape of the PSF across the field of view. 
In this work, we use realistic, altitude-dependent weather and turbulence input provided by the \psfws package (\figref{inputs}) to simulate the atmospheric PSF across the Rubin Observatory field of view. 
We summarize our findings here:
\begin{enumerate}
\item 
The variance in PSF size and shape across a single exposure decreases as wind speed increases (see \figref{paramvar}).
\item 
The values of the 2PCFs for PSF shape in a single exposure are of the same order as (or larger than) the expected 2PCFs for cosmic shear over the range of angular separations used in cosmic shear analyses: from a few arcmin to over 100\amin 
(see \figref{1d2pcf}). 
\item 
There exist dominant wind directions at \cp (red histograms in top panels in \figref{bigscatter}), which in turn lead to dominant orientations of anisotropies in the 2PCF $\xi_+$ (red histograms in right panels in \figref{bigscatter}). 
At scales of $\sim100\amin$, these anisotropies are due to strong correlations with ground-layer wind direction (upper left  scatter plot in \figref{bigscatter}); at scales of only a few arcmin, they are due to correlations with free-atmosphere wind direction (lower right scatter plot). 
As discussed in \secref{inputcompare}, these results can be understood in terms of the different angles subtended at different heights by turbulent structure of the same physical scale. 
\end{enumerate}

PSF modeling and interpolation methods must accurately capture the anisotropic two-point correlations for PSF size and shape on different scales. 
In the future, high fidelity simulations generated with \psfws can be used to test whether current modeling and interpolation methods \citep[e.g., those implemented in][]{jarvis_dark_2020} can reach the necessary accuracy. 
One interpolation technique that merits further exploration is anisotropic Gaussian process interpolation; see, for example, \cite{leget_improving_2021} and \cite{fortino_reducing_2021} for applications to PSF astrometry. 

Because of the dominant wind direction at \cp, we expect to see a dominant orientation of the anisotropy in the 2PCF for PSF size and shape with respect to the ground coordinate system, across LSST exposures at Rubin Observatory.  
The mapping of wind direction on the ground onto sky coordinates is determined by the pointing of the telescope for each exposure. 
Therefore, the degree to which the dominant wind direction will vary on the sky for a single field depends on the observing strategy for the survey. 
Using the observing strategy for the 300-square-degree DESC DC2 simulation, we find that the dominant wind direction from \figref{bigscatter} will translate to a persistent on-sky anisotropy; further study is needed to understand the implications for the full LSST survey. 
PSF simulations produced with \psfws input can be used to study this question for a particular survey strategy. 

The \psfws software package for producing correlated weather and turbulence input to simulations, configurable to any observatory, is public at \url{https://github.com/LSSTDESC/psf-weather-station} and includes installation instructions, documentation, and tutorial notebooks.

\section*{Acknowledgements} 
This paper has undergone internal review in the LSST Dark Energy Science Collaboration by Gary Bernstein, Mike Jarvis, and Arun Kannawadi; we thank them for their constructive comments and reviews. 
We thank Mike Jarvis, Arun Kannawadi, and Morgan Schmidt for their code review of the \psfws package, and Mike Jarvis, Sowmya Kamath, Pierre-François L\'eget, and Sidney Mau for useful discussions.
We thank James Osborn for sharing his expertise.
C-AH and PRB are supported in part by Department of Energy Office of Science grant DE-SC0009841.
C-AH acknowledges support from the DOE Computational Science Graduate Fellowship Program (DE-FG02-97ER25308) and the Stanford University DARE Doctoral Fellowship Program. 
The DESC acknowledges ongoing support from the Institut National de Physique Nucl\'eaire et de Physique des Particules in France; the Science \& Technology Facilities Council in the United Kingdom; and the Department of Energy, the National Science Foundation, and the LSST  Corporation in the United States.  
DESC uses resources of the IN2P3  Computing Center (CC-IN2P3--Lyon/Villeurbanne - France) funded by the Centre National de la Recherche Scientifique; the National Energy Research Scientific Computing Center, a DOE Office of Science User Facility supported by the Office of Science of the U.S.\ Department of Energy under Contract No.\ DE-AC02-05CH11231; STFC DiRAC HPC Facilities, funded by UK BEIS National E-infrastructure capital grants; and the UK particle physics grid, supported by the GridPP Collaboration.  
This work was performed in part under DOE Contract DE-AC02-76SF00515.
Generated using Copernicus Climate Change Service information (2019). 
We acknowledge ECMWF for access to the ERA5 data through the MARS access system.
The computing for this project was performed on the Stanford Sherlock cluster.
We would like to thank Stanford University and the Stanford Research Computing Center for providing computational resources and support. 

\section*{Author Contributions}

Claire-Alice H\'ebert developed the \psfws software package, performed the main analysis, and produced all the plots in the paper.
Patricia Burchat contributed to defining the project, and advised throughout.
C-AH and PB contributed equally to writing the paper. 
Joshua Meyers advised on all aspects of the project and reviewed the \psfws package.
My H.~Do contributed to the analysis early in the project as a participant in the Cal-Bridge program for undergraduate students.

\appendix
\section*{Global Circulation Models}\label{app:gcm}
As summarized in \secref{inputs}, multiple organizations around the world produce high-quality weather models and forecasts; we focus here on those from the European Centre for Medium-Range Weather Forecasts (ECMWF) and the National Oceanic and Atmospheric Administration National Centers for Environmental Prediction (NOAA NCEP). 
Data products from these global models can be very useful for studies of the atmosphere for astronomical applications.
Here we summarize the types of data available and considerations for their use in atmospheric PSF simulations.

Both ECMWF and NCEP make available two types of data: 
(1) \textit{analysis} products are the best estimate of the state of the atmosphere, produced by combining a numerical weather prediction model with a variety of observations through a process called data assimilation; 
(2) \textit{forecast} products are the numerical predictions (based on initial analysis products) for some time into the future.
Analysis and forecast data are available in real-time (of use for weather forecasting) and as \textit{reanalysis} data products: state-of-the-art data assimilation and numerical modeling methods applied to archival data (highly relevant for long-term climate monitoring).\footnote{All available atmospheric reanalysis datasets are summarized at \url{https://reanalyses.org/atmosphere/comparison-table}}

The 5th generation ECMWF reanalaysis (ERA5) catalog covers the time period from 1940 to the present and is extensively documented\footnote{ERA5: \url{https://confluence.ecmwf.int/display/CKB/ERA5}} \citep{hersbach_era5_2020}.
ERA5 analyses are available hourly, with forecasts initialized at 00h and 18h UTC.
At the time of writing, all ECMWF archival data (including ERA5) and subsets of real-time forecasts are available publicly under creative commons.\footnote{ECMWF licenses: \url{https://www.ecmwf.int/en/forecasts/accessing-forecasts/licences-available}}

All NOAA NCEP data are available publicly; real-time forecasts are from the Global Forecast System (GFS), and several reanalysis efforts are available for specific time periods.\footnote{NOAA GFS: \url{https://psl.noaa.gov/data/gridded/data.ncep.reanalysis.html}}
The NCEP analysis products are available every 6h.

Data products are output on a uniform grid over the Earth's surface, with resolution 31\km $\approx 0.28^\circ$ for ERA5 and NCEP GFS, and $2.5^\circ$ for NCEP reanalysis.

Data products are available at heights corresponding to specific levels of pressure rather than specific altitudes.
Since higher spatial resolution in the vertical direction allows for more accurate capture of important wind gradients in the atmosphere, we use the output type with densest vertical coverage -- called \textit{model levels}. 
ECMWF uses 137 model levels. 
NCEP uses 127 for the period since February 3, 2021, and 64 prior to that time.
Model levels follow terrain at the Earth's surface, and the conversion to altitude uses temperature and specific humidity.
This conversion has been implemented in \psfws following ECWMF documentation and example code found in the Q\&A section of the ERA5 wiki.

We choose to use ECMWF data products because the temporal and spatial resolution of available data is higher than for NCEP.
In addition, ECMWF documentation is more detailed and accessible.

\bibliography{references}{}
\bibliographystyle{aasjournal}

\end{document}